\documentclass[lettersize,journal]{IEEEtran}
\usepackage{algorithmic}
\usepackage{algorithm}
\usepackage{array}
\usepackage{breqn} 
\usepackage[caption=false,font=normalsize,labelfont=sf,textfont=sf]{subfig}
\usepackage{textcomp}
\usepackage{stfloats}
\usepackage{url}
\usepackage{verbatim}
\usepackage{graphicx}
\usepackage{cite}
\usepackage{amsmath,amssymb,amsfonts}
\usepackage{algorithmic}
\usepackage{graphicx}
\usepackage{textcomp}
\usepackage{xcolor}
\usepackage{array}
\usepackage{mdwmath}
\usepackage{mdwtab}
\usepackage{eqparbox}
\usepackage{url}
\usepackage{braket}
\usepackage{cite}
\usepackage{doi}
\usepackage{hyperref}
\usepackage{multirow}
\usepackage{amssymb}
\usepackage{braket}
\usepackage{amsmath}
\begin{document}

\title{Quantum dot-based high-fidelity universal quantum gates in noisy environment}

\author{Yash Tiwari, Aditya Dev,
      Vishvendra Singh Poonia
\thanks{This work is supported by the Science and Engineering Research Board, Department of Science and Technology (DST), India, with Grants No. CRG/2021/007060 and No. DST/INSPIRE/04/2018/000023.}
 \thanks{Vishvendra Singh Poonia and Yash Tiwari are with the Department of Electronics and Communication Engineering, Indian Institute of Technology, Roorkee, India. (Email: vishvendra@ece.iitr.ac.in)}
  \thanks{Aditya Dev is with the Department of Physical Sciences, Indian Institute of Science Education and Research (IISER) Mohali.}
}

\markboth{IEEE Transactions on Nanotechnology 2023}%
{Shell \MakeLowercase{\textit{et al.}}: A Sample Article Using IEEEtran.cls for IEEE Journals}


\maketitle

\begin{abstract} 
Quantum dot-based spin qubit realization is one of the most promising quantum computing systems owing to its integrability with classical computation hardware and its versatility in realizing qubits and quantum gates. In this work, we investigate a quantum dot-based universal set of quantum gates (single qubit gates and the Toffoli gate) in the presence of hyperfine fluctuation noise and phononic charge noise. We model the spin dynamics and noise processes in the NOT gate, Hadamard gate and the Toffoli gate using the Lindblad master equation formalism to estimate the operating ranges of the external static and ac magnetic fields to achieve high fidelity operation of these gates in a noisy environment. In addition, the generality of the framework proposed in this paper enables modeling of larger quantum processors based on spin qubits in realistic conditions.
\end{abstract}

\begin{IEEEkeywords} 
Quantum dots, spin qubits, dephasing, universal quantum gates, Hadamard gate, Toffoli gate
\vspace{-1em}
\end{IEEEkeywords}

\section{Introduction}
In 1998, Loss and DiVincenzo proposed a quantum dot based implementation of quantum computation that used electron spin as a qubit~\cite{loss1998quantum}. Since then, tremendous progress has been made on quantum dot based spin qubits~\cite{koppens2006driven,maune2012coherent,veldhorst2014addressable,veldhorst2015two,zajac2018resonantly,russ2018high,zajac2015reconfigurable,petta2005coherent,hanson2007spins,zajac2018single}. To implement an arbitrary quantum operation, we need to have a set of `universal' quantum gates. There exist many such sets that can act as universal quantum gates according to Solovay-Kitaev theorem~\cite{nielsen2001quantum,harrow2002efficient}. However, qubits interact with the environment, which affects the gate implementation and deteriorates their fidelity. For example, in a GaAs QD system, we observe interactions with i) spin of surrounding nuclei and ii) phonons~\cite{taylor2006hyperfine,amasha2008electrical,khaetskii2001spin,san2006geometrical,marquardt2005spin,borhani2006spin,kouwenhoven1997electron,simmons2007single}. Similar is the case with Si quantum dot based qubits. Both of these are the primary sources of decoherence in the qubits. The spin of the nuclei surrounding the confined electron in the quantum dot exhibits a normalized magnetic field distribution. This magnetic field interacts with the spin of a confined electron introducing hyperfine interaction (HFI) noise in the system. The interaction with phonons essentially captures the charge noise due to lattice vibrations. It is mediated by the spin-orbit interaction (SOI)~\cite{camenzind2018hyperfine,alcalde2008spin}. Generally, spin qubits cannot be directly affected by electric field fluctuations. Spin-orbit interaction (SOI) mixes the spin states with the orbital (electron cloud) state making the system susceptible to electrostatic fluctuations~\cite{amasha2008electrical}.  As pointed out earlier as well, the phonon interaction is the dominant source of charge fluctuation in the quantum dot environment that we have taken in our study. It has been observed that HFI noise is dominant at the low external static magnetic field (low energy gap), and phonon noise is dominant at the high value of static magnetic field (high energy gap)~\cite{amasha2008electrical}. The silicon-based spin qubits are relatively free from hyperfine interactions and therefore they are well suited for spin-based quantum computing. However, they suffer from a problem inherent to their electronic structure. The conduction band minima of silicon have six-fold degeneracy providing an uncontrolled degree of freedom for electrons that are used to define qubits and it may cause relaxation and dephasing in the qubit state~\cite{zwanenburg2013silicon,eriksson2013semiconductor}. 

In this work, we present a general framework to model the spin dynamics of the universal set of quantum gates in noisy environment. We examine the fidelity of the gates and find out the parameter regime to achieve the high-fidelity ($\geq$ 98\%) gate operations. 
One of the most popular universal gate sets is the combination of Toffoli gate and Hadamard gate~\cite{aharonov2003simple,vilmart2022completeness}. We identify a common parameter regime where both gate operations show high fidelity. The proposed framework is general in nature and can be extended to more qubits and larger quantum processors based on spin qubits. This would also allow one to include quantum control techniques used to enhance the gate fidelity further and reduce the error correction overhead. In this framework, we use Lindblad master equation formalism to model the spin dynamics and decoherence due to noise in the spin system. 

The manuscript has been organized as follows: Section~\ref{Methodology} discusses the modeling and numerical methodology followed for the analysis. Section~\ref{Single_Qubit} discusses the result for single qubit quantum gates. In addition to the Hadamard gate operation, we also discuss NOT gate in detail in this section. Section~\ref{Three_Qubit} discusses the results for the three-qubit Toffol gate. Finally, we conclude the paper with a discussion about extending the framework and identifying the future directions. 

\section{Methodology}
\label{Methodology}
In this section, we discuss the modeling of qubit spin dynamics and noise processes therein. We use the Lindblad operator equation for this purpose that can be written as (Eq.~\ref{Master_Eqaution}):
\begin{equation}\label{Master_Eqaution}
 \frac{d\rho(t)}{dt}= L_0 + L_D.
\end{equation}
\begin{equation}\label{Master_Eqaution_Cohrent_Part}
 L_0= -\frac{i[H(t), \rho(t)]}{\hbar}.
\end{equation}
A quantum system’s coherent and non-coherent time evolution is observed due to $L_0$ and $L_D$, respectively. The general
formulation of $L_0$ is given in Eq.~\ref{Master_Eqaution_Cohrent_Part}, where $H(t)$ is the complete Hamiltonian of a closed system, $\rho(t)$ denotes quantum state at arbitrary time instant t. The non-coherent evolution operator $L_D$  is 
 described in Eq.~\ref{Non_cohrent_master1}.
\begin{equation}\label{Non_cohrent_master1}
 L_{D}=\sum_{n}\frac{1}{2}\{ 2C_{n}\rho(t)C_{n}^\dagger-\rho(t)C_{n}^\dagger C_{n}-C_{n}^\dagger C_{n}\rho(t)\}
\end{equation}
The $C_n$, in general, would be associated with a rate and an accompanying operator. $C_n$ due to relaxation would have an operator of form $\ket{w_j}\bra{w_k}$ where the energy relaxation would be happening from energy level $w_k$ to $w_j$. The rate associated with the operator would be a function of the energy gap($E_{wj}$-$E_{wk}$). Operator associated due to dephasing is of the form of $\sigma_z$, where $\sigma_z$ is the Pauli-Z matrix. The rate associated with the dephasing is the inverse of $T_2^*$ which is a material dependent parameter. In this work, we have used parameters corresponding to GaAs quantum dot system, however, the framework presented is quite general and applicable to other spin qubit systems including the ones based on Si quantum dots.

\begin{equation}\label{Non_cohrent_master2}
\begin{split}
 C_{i1} = {\sqrt{\Upsilon^+_{jk}}}\ket{w_j}\bra{w_k} =
\sqrt{{\Upsilon}e^{\frac{-{\omega_{jk}^{2}}}{2\delta E_{nuc}^{2}}}}\ket{w_j}\bra{w_k}
\end{split}
\end{equation}
\begin{equation}\label{Non_cohrent_master3}
\begin{split}
 C_{i2} = {\sqrt{\Upsilon^-_{kj}}}\ket{w_k}\bra{w_j} =
\sqrt{{\Upsilon}e^{\frac{-{\omega_{kj}^{2}}}{2\delta E_{nuc}^{2}}+\frac{\omega_{kj}}{T_{K}})}}\ket{w_k}\bra{w_j}
\end{split}
\end{equation}
\begin{equation}\label{Non_cohrent_master4}
\begin{split}
 C_{i3} = {\sqrt{P^{+}_{jk}}}\ket{w_j}\bra{w_k}=\sqrt{P \Bigg| \frac{\omega_{jk}^3 E_{jk}^{2}} {1-e^{-\frac{\omega_{jk}}{T_k}}} \Bigg|}\ket{w_j}\bra{w_k}
 \end{split}
\end{equation}
\begin{equation}\label{Non_cohrent_master5}
\begin{split}
 C_{i4} = {\sqrt{P^{-}_{kj}}}\ket{w_k}\bra{w_j}=\sqrt{P \Bigg| \frac{\omega_{kj}^3 E_{kj}^{2}} {1-e^{-\frac{\omega_{kj}}{T_k}}} \Bigg|}\ket{w_k}\bra{w_j}
 \end{split}
\end{equation}
Eq.~\ref{Non_cohrent_master2} to Eq.~\ref{Non_cohrent_master5} denote the relaxation operators in a spin qubit system. 
In Eq.~\ref{Non_cohrent_master2} and Eq.~\ref{Non_cohrent_master4}, $w_{jk}>0$ corresponds to positive transition (energy absorption) whereas in Eq.~\ref{Non_cohrent_master3} and Eq.~\ref{Non_cohrent_master5}, $w_{kj}<0$ corresponds to negative transition (energy emission).
$C_{i1}$ ($\Upsilon^+_{jk}$) and $C_{i2}$ ($\Upsilon^-_{kj}$) are positive and negative relaxation operators (rates) due to HFI noise. Similarly, $C_{i3}$ ($P^+_{jk}$) and $C_{i4}$ ($P^-_{kj}$) are positive and negative relaxation operators (rates) due to phononic noise \cite{mehl2013noise}. 
In $C_{i1}$ and $C_{i2}$, $\delta E_{nuc}$ = 0.3 $\mu$eV for GaAs~\cite{taylor2006hyperfine,paget1977low}.  The parameter $\delta E_{nuc}$ is associated with the nuclear magnetic field (due to nuclear spin) experienced by the confined electron. $\Upsilon$ is the constant associated with relaxation due to hyperfine noise, in $C_{i1}$ and $C_{i2}$.
The temperature of the system is $T_{k}$ = 10 $\mu$eV (125mK).
$P$ is the constant associated with the relaxation operator due to phononic interactions, whereas $E_{kj}$ corresponds to Zeeman split between the energy levels ~\cite{mehl2013noise,amasha2008electrical}. The dephasing operator is given in Eq.~\ref{Dephasing_operator_single_spin} where the value of $T_2^*$ is taken from an experiment~\cite{koppens2008spin}.

\begin{equation}\label{Dephasing_operator_single_spin}
  C_5=L_{Dephasing}=\sqrt{\frac{1}{2T_{2}^*}}\sigma_z 
\end{equation}

\begin{figure}[htbp]
\centering
\includegraphics[width=2.5in]{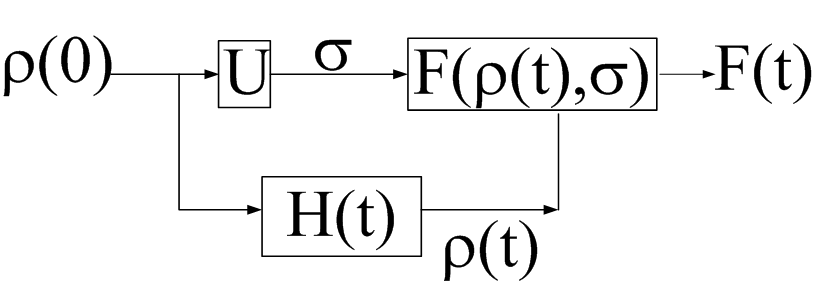}
\vspace{-0.5em}
\caption{Schematic used to calculate fidelity of quantum gates. $U$ is the matrix corresponding to the gate operation, $H(t)$ is the system Hamiltonian. $\rho(t)$ is the state of the system and $\sigma$ is the output after multiplication with gate matrix $U$. F(t) is the state fidelity between $\rho$ and $\sigma$.}
\label{Fied_Schematic}
\end{figure}

In order to assess the efficacy of the gate operations, we use `state fidelity' as a quantifier. State fidelity between two states $\rho$ and $\sigma$ is defined in Eq.~\ref{Fidelity} where $Tr$ is the trace of a matrix. We use the schematic shown in Fig.~\ref{Fied_Schematic} to calculate the fidelity of a gate operation as a function of time where $U$ is the gate matrix, $H(t)$ is the system hamiltonian, $\rho(t)$ is the state of the system and $\sigma$ is output after multiplication with gate matrix $U$. $\sigma$ remains constant with time. State fidelity also helps us to understand the effect of relaxation and dephasing on a gate operation.
\begin{equation}\label{Fidelity}
\begin{split}
F(\rho,\sigma) = \left(Tr \left( \sqrt{\sqrt{\rho}\sigma\sqrt{\rho}} \right) \right)^2
 \end{split}
\end{equation}

In summary, the evolution of the qubit states and gate operation is modeled by the Lindblad master equation, and the fidelity of gate operations is evaluated through the schematic in Fig.~\ref{Fied_Schematic}. This analysis leads us to estimate the values of external parameters (dc and ac magnetic field values) to achieve high-fidelity gate operations.
We employ this methodology on NOT gate, Hadamard gate, and Toffoli gate operations. We also study the effect of noise on the operations of all these gates.

\section{Single qubit Quantum Gates}\label{Single_Qubit}
The Hamiltonian of single-qubit gate operation (i.e. electron spin resonance (ESR)) is given by Eq.~\ref{Hamiltonian}:
\begin{equation}\label{Hamiltonian}
 H(t) = g\mu B_{static}\sigma_z - g\mu B_{ac}(\cos(\omega t )\sigma_x - \sin(\omega t )\sigma_y)
\end{equation}
where $g$ is the gyromagnetic ratio, $B_{static}$ and $B_{ac}$ are the static and ac magnetic fields applied to the system. $\sigma_x, \sigma_y, \sigma_z$ are the Pauli matrices. When the initial state of the system can be defined as: $\ket{\Phi(0)} = {a_0} \ket{\uparrow} + {b_0}   \ket{\downarrow}$,
$\ket{\Phi(t)}$ is given as: $\ket{\Phi(t)}={a} \ket{\uparrow} + {b}   \ket{\downarrow}$ where 

\begin{dmath}
  a = \left[ a_0 \left( \cos \left( \frac{\omega_1 t}{2f}\right) - i f \frac{w - \omega_0}{\omega_1}\sin \left( \frac{\omega_1 t}{2f}\right)\right)
    + b_0 i f \sin \left( \frac{\omega_1 t}{2f}\right) \right] e^{i w t / 2}
\end{dmath}
\begin{dmath}
  b = \left[ b_0 \left( \cos \left( \frac{\omega_1 t}{2f}\right) - i f \frac{w - \omega_0}{\omega_1}\sin \left( \frac{\omega_1 t}{2f}\right)\right)
    + a_0 i f \sin \left( \frac{\omega_1 t}{2f}\right) \right] e^{-i w t / 2}
\end{dmath}
$\omega_0 = g\mu B_{static}$, $\omega_1 = g\mu B_{ac}$ and $\omega$ is the applied frequency of ac magnetic field. $f$ is defined by Eq.~\ref{f_exp} as:

\begin{equation}\label{f_exp}
f = \sqrt{\frac{\omega_1 ^2}{\left(w - \omega_0\right)^2 + \omega_1}}
\end{equation}
 The value of $f$ is unity at resonance, i.e. $\omega=\omega_0$~\cite{FAMU_FSu,le2006short} .
When initial state is: $\ket{\Phi(0)}=\ket{\uparrow}$ ($a_0=1$ and $b_0=0$), $\Phi(t)$ will be given as:
\begin{equation}\label{General_Eq}
    \ket{\Phi(t)} = \left[ 
    \cos \left (\frac{\omega_1 t}{2}\right)e^{i \omega_0 t / 2} \ket{\uparrow} + i \sin\left(\frac{\omega_1 t}{2}\right)e^{-i \omega_0 t / 2} \ket{\downarrow}
    \right]
\end{equation}

With this background, in two subsections, we will discuss the implementations of single qubit NOT and Hadamard gates.
\subsection{NOT gate implementation}
\begin{figure}[htbp]
\centering
\includegraphics[width=90mm]{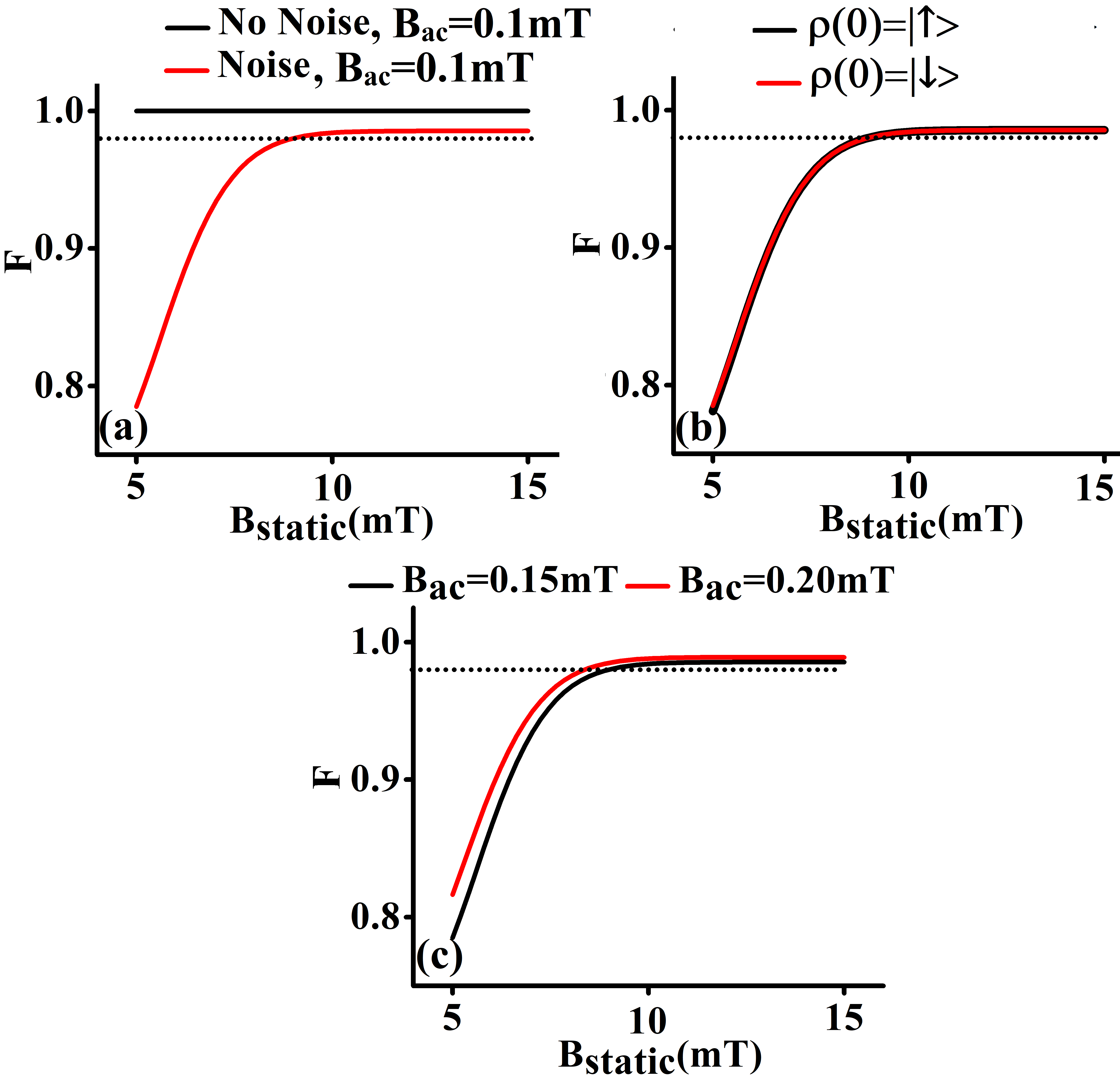}

\caption{(a) Fidelity vs $B_{static}$ for NOT gate operation at $B_{ac}$ = 0.1mT, when there is no noise (black), noise (red) in the system. (b) Fidelity comparison for NOT gate when $\rho(0)=\ket{\uparrow}$ (black) and $\rho(0)=\ket{\downarrow}$ (red) at $B_{ac}$ = 0.2mT  (c) Fidelity vs $B_{static}$  for two values of $B_{ac}$ = 0.15mT (black), 0.20mT (red). At high $B_{ac}$ a much greater range of $B_{static}$ can  be used corresponding to F $\geq$ 0.98 for NOT gate operation.}
\label{NOT_Low}
\end{figure}

The NOT gate is implemented according to the scheme shown in Fig.\ref{Fied_Schematic} where $U = [[0,1],[1,0]]$ and $H(t)$ would be same as Eq.~\ref{Hamiltonian}. The $F(\rho(t),\sigma=\ket{\downarrow})$ when $\ket{\Phi(0)}=\ket{\uparrow}$ is $\sin^2\frac{\omega_1 t}{2}$ at resonant frequency $\omega=\omega_0$. We achieve $F=1$ when $\frac{\omega_1 t}{2}=\frac{\pi}{2}$ at which NOT gate operation occurs. In Fig.~\ref{NOT_Low}(a), we have shown the fidelity ($F$) for a single qubit NOT operation as a function of $B_{static}$ at low values of static magnetic field (few mT). The black (red) curve corresponds to the case when no noise (noise) is considered in the system. We take a threshold of 98\% fidelity to find out the parameter regime where gate operation is achieved with high fidelity ($>$ 98\%). The dotted lines correspond to $F=0.98$. At values of $B_{static}\leq7mT$, we observe that $F\leq0.98$ due to hyperfine noise. In Fig.\ref{NOT_Low}(b), we observe that whether initially $\rho(0)=\ket{\uparrow}$ or $\rho(0)=\ket{\downarrow}$, the system show drop in $F$ in a similar manner due to noise.

\begin{figure}[htbp]
\centering
\includegraphics[width=90mm]{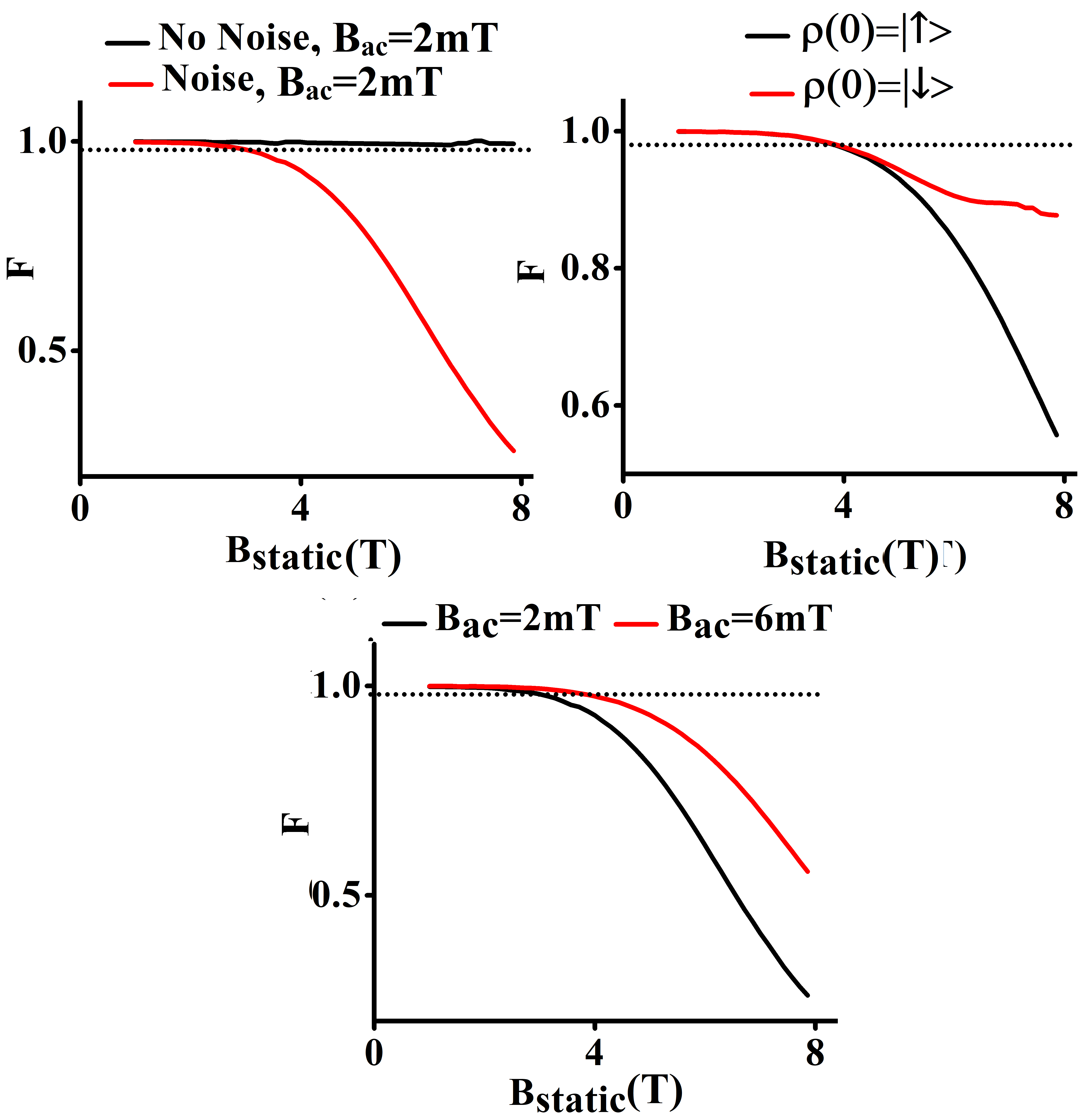}

\caption{(a) Fidelity vs $B_{static}$ for NOT gate operation at $B_{ac}$ = 2mT, when there is no noise (black), noise (red) in the system. (b) Fidelity comparison for NOT gate when $\rho(0) = \ket{\uparrow}$ (black) and $\rho(0)=\ket{\downarrow}$ (red) at $B_{ac}$ = 6mT  (c)Fidelity vs $B_{static}$  for two values of $B_{ac}$  =2mT (black), 6mT (red). At high $B_{ac}$ a much greater range of $B_{static}$ can be used corresponding to F $\geq$ 0.98 for NOT gate operation.}
\label{NOT_High}
\end{figure}
 In Fig.\ref{NOT_Low}(c), we observe that as the ac field ($B_{ac}$) increases, $B_{static}$ at which $F\geq0.98$ decreases, thus increasing the working range of $B_{static}$. We also define a quantity called $B_F$ which is the $B_{static}$ at which system $F=0.98$ with noise. A curve correlating $B_F$ with $B_{ac}$ has been examined in Ref.~\cite{tiwari2022effect}. 

 In Fig.\ref{NOT_High}(a), we show $F$ for a single qubit NOT gate operation as a function of $B_{static}$ at high values of static magnetic field (few Tesla). The black (red) curve corresponds to the case when no noise (noise) is considered in the system. At values of $B_{static}\geq 3T$, we observe a fall in $F$ due to phononic noise. In Fig.~\ref{NOT_High}(b), we observe that till $F=0.98$ threshold, $\rho(0)=\ket{\uparrow}$ and $\rho(0)=\ket{\downarrow}$ behave similarly. However, at extremely high $B_{static}$ and with $\rho(0)=\ket{\downarrow}$, the system $F$ decays slower compared to the case when $\rho(0)=\ket{\uparrow}$ due to noise. In Fig.\ref{NOT_High}(c), as the ac field $B_{ac}$ increases, $B_F$ also increases. Thus increasing the working range of $B_{static}$.

\subsection {Hadamard gate implementation}
The Hadamard gate is implemented according to Fig.~\ref{Fied_Schematic} with $U = \frac{1}{\sqrt{2}} [[1,1],[1-1]]$ and $H$ would be same as Eq.~\ref{Hamiltonian}. The $F(\rho(t),\sigma=\frac{1}{\sqrt{2}}(\ket{\uparrow}+\ket{\downarrow})$ when $\ket{\Phi(0)}=\ket{\uparrow}$ is $1+\sin(\omega_0t)\sin(\omega_1t)$ at resonant frequency $\omega=\omega_0$. We achieve $F=1$ when either $\omega_0t=n\pi$ or $\omega_1t=m\pi$ or both (where $n ,m$ are integers). At this instance, the Hadamard gate operation takes place.

\begin{figure}[htbp]
\centering
\includegraphics[width=90mm]{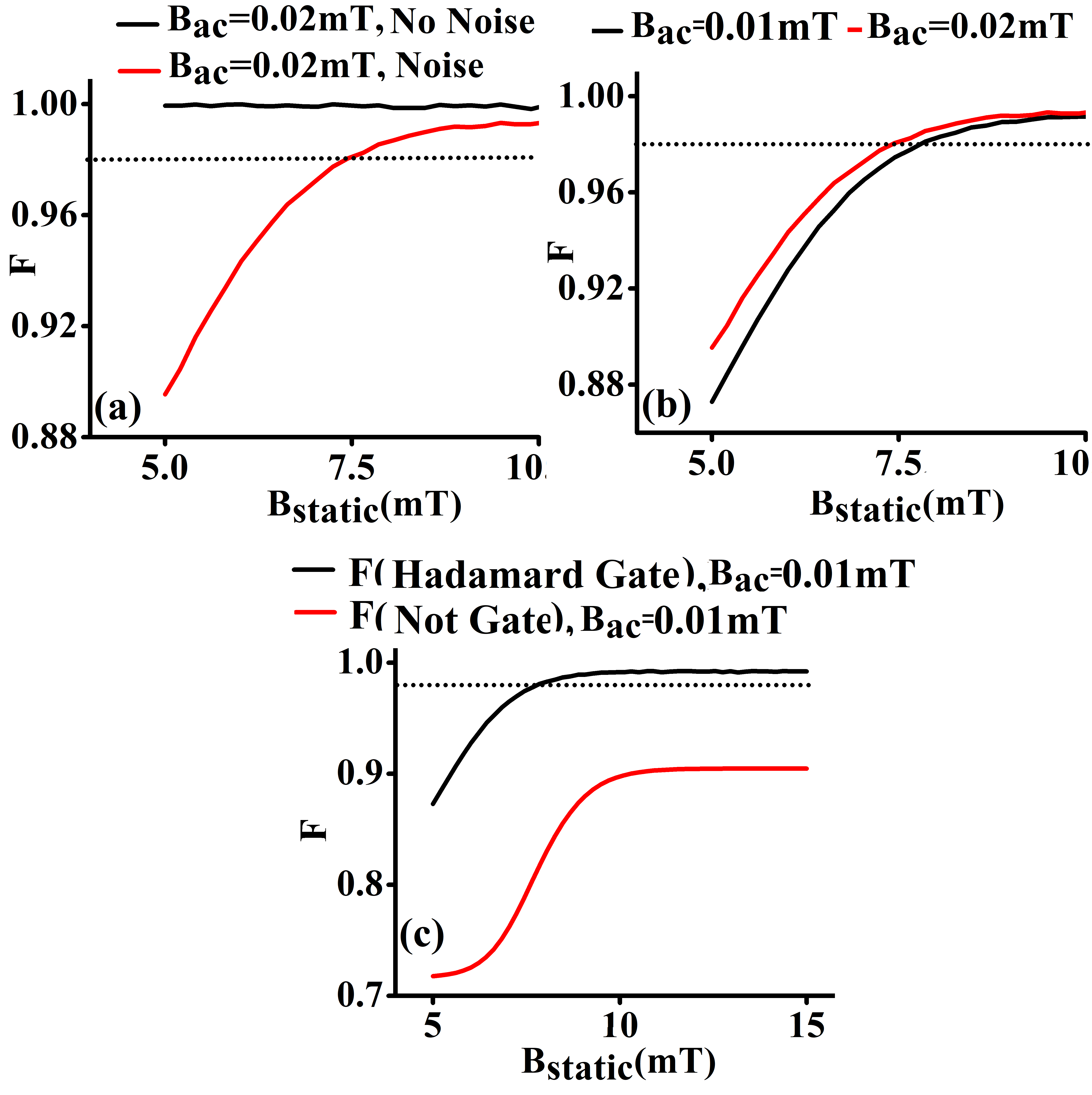}

\caption{(a) Fidelity vs $B_{static}$ for Hadamard gate at $B_{ac}$ = 0.02mT, when there is no noise (black), noise (red) in the system. (b) Fidelity vs $B_{static}$  for two values of $B_{ac}$ = 0.01mT (black), 0.02mT (red). At high $B_{ac}$ a much greater range of $B_{static}$ can  be used corresponding to F $\geq$ 0.98 for Hadamard gate operation. (c) Fidelity comparison for NOT gate and Hadamard gate operation at $B_{ac}$ = 0.01mT.}
\label{Hadmard_low}
\end{figure}

In Fig.~\ref{Hadmard_low}(a), we have shown $F$ for a single qubit Hadamard gate operation as a function of $B_{static}$ at low values of static magnetic field (few mT). The black (red) curve corresponds to the case when no noise (noise) is considered in the system. The dotted lines correspond to $F=0.98$. Both cases are taken at $B_{ac}$ = 0.02mT. At low values of $B_{static}$, we observe a drop in $F$ due to hyperfine interaction noise. In Fig.~\ref{Hadmard_low}(b), as the ac field $B_{ac}$ is increased, $B_{static}$ at which $F\geq0.98$($B_F$) decreases, thus, increasing the working range of $B_{static}$. In Fig.~\ref{Hadmard_low}(c), we illustrate two curves highlighting the differences in fidelity for the NOT gate and the Hadamard gate for the same value of $B_{ac}$ =0.01 mT. It appears that at a low value of $B_{ac}$, Hadamard gate exhibits $F\geq0.98$ whereas NOT gate does not show high fidelity ($F\geq0.98$) and requires a higher ac magnetic field for its operation. 
\begin{figure}[htbp]
\centering
\includegraphics[width=90mm]{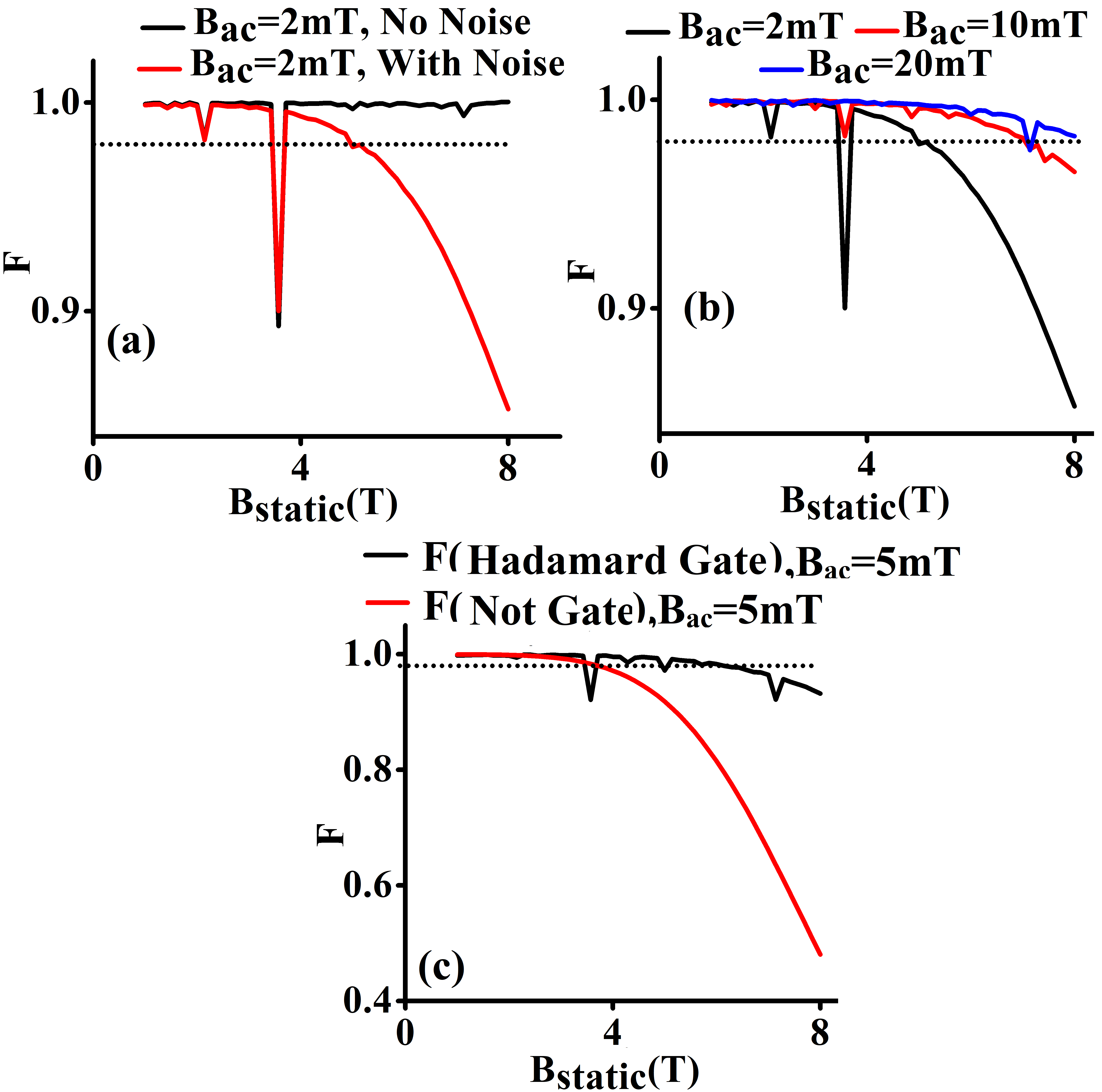}

\caption{(a) Fidelity vs $B_{static}$ for hadamard gate at $B_{ac}$ = 2mT, when there is no noise (black), noise (red) in the system. The peak at $B_{static}$ = 3.5T is  due to $B_{ac}$ and $B_{static}$ values and is not due to decoherence. (b) Fidelity vs $B_{static}$  for three values of $B_{ac}$ = 2mT (black), 10mT (red) and 20mT (blue). This shows that the fidelity drop at $B_{static}$=3.5T is removed at high values of $B_{ac}$. Moreover, at high $B_{ac}$ a much greater range of $B_{static}$ can  be used corresponding to F $\geq$ 0.98 for hadamard gate operation. (c) Fidelity comparison for NOT gate and hadamard gate operation at $B_{ac}$ = 5mT  }
\label{Hadmard_High}
\end{figure}

In Fig.~\ref{Hadmard_High}(a), we show $F$ for a single qubit Hadamard gate operation as a function of $B_{static}$ at high values of static magnetic field (few Tesla). The black (red) curve corresponds to the case when no noise (noise) is considered in the system. Both cases are taken at $B_{ac}$ = 0.2 mT. We found a sharp drop in $F$ for both with and without noise cases at $B_{static}=3.5T$. This suggests that the drop in $F$ is not related to the decoherence but it is due to relative values of $B_{static}$ and $B_{ac}$. At high values of $B_{static}\geq$5T, we observe a drop in $F$ due to hyperfine interaction noise. In Fig.~\ref{Hadmard_High}(b), as the ac field $B_{ac}$ is increased, $B_{static}$ at which $F\geq0.98$ also increases. Thus, increasing the working range of $B_{static}$. We also observe that the sharp dip in $F$ at $B_{static}=3.5T$ is reduced at high $B_{ac}$ giving an extra advantage from the point of fidelity. In Fig.~\ref{Hadmard_High}(c), we illustrate two curves highlighting the difference in fidelity for NOT gate and Hadamard gate operations for the same value of $B_{ac}$ = 5mT. It appears that Hadamard gate shows $F\geq0.98$ for a greater range of $B_{static}$ compared to the NOT gate.

\subsection{Relation between $T_2^*$ and $B_{ac}$}
\begin{figure}[htbp]
\centering
\includegraphics[width=90mm]{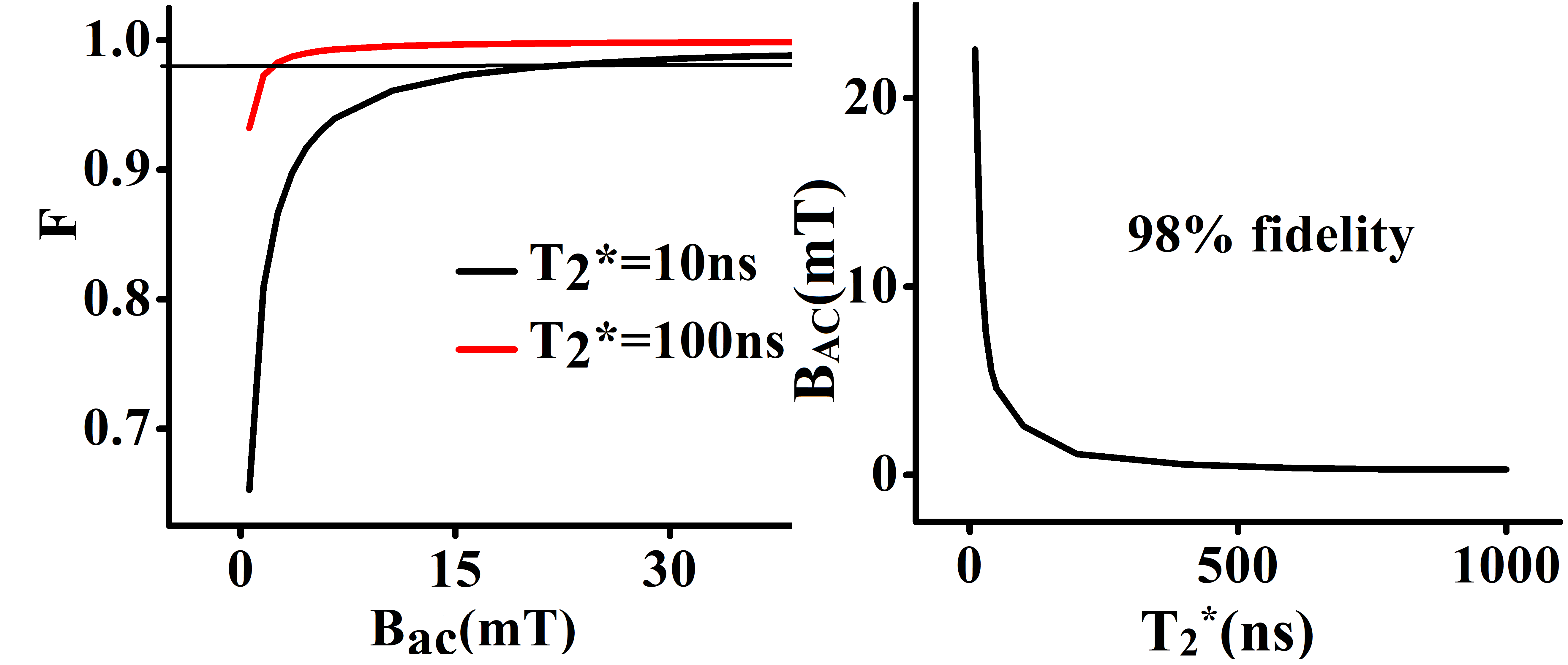}
\caption{(a) Fidelity vs $B_{ac}$ for cases $\rho(0)=\ket{\uparrow}$ at two different $T_2^{*} =$ 10ns (black), 100ns (red) at $B_{static}$ = 1T for NOT gate. The horizontal line corresponds to $F$ = 0.98 and at that point $B_{ac}=B_{AC}$. All other parameters are kept constant. (b) $B_{AC}~$vs$~T_2^{*} $ depicting that at higher $T_2^*$, gate operation require lower $B_{AC}$ for high fidelity ($\geq$0.98) NOT gate operation. }
\label{Fig2}
\end{figure}

In this subsection, we examine the relation between the ac magnetic field and gate fidelity for different values of dephasing time. We also analyze the value of ac magnetic field required to achieve high fidelity operation for a wide range of dephasing time values. Fig.~\ref{Fig2} shows the relation of $B_{ac}$ and $T_2^*$. The $T_2^*$ is correlated with the dephasing time mentioned in Eq.~\ref{Dephasing_operator_single_spin}. In Fig~\ref{Fig2}(a), we have plotted $F$ for NOT gate operation when $\rho(0)=\ket{\uparrow}$ at $B_{static} = 1T$. This has been done for two values of $T_2^*$. The dotted line corresponds to $F$=0.98. We define a quantity called $B_{AC}$ which is the ac magnetic field when $F$=0.98 at a constant value of $B_{static}$. In Fig~\ref{Fig2}(b) we plot values of $B_{AC}$ for different values of $T_2^*$ which is required for the high fidelity gate operation. The relation shows an exponential decay highlighting that when $T_2^*$ is relatively low (few $ns$) a higher $B_{AC}$ is required. As $T_2^*$ increases, $B_{AC}$ is reduced and eventually it saturates and becomes independent of $T_2^*$.

In summary, we observe that a higher value of ac field mitigates decoherence at both high and low static magnetic field regimes. This is because raising the value of the ac magnetic field decreases the gate operation time (faster gate operation). Hence, the system gets much less time to interact with its environment, leading to lesser decoherence and better performance at both low and high static magnetic field values. We also observe that Hamdard gate implementation shows a much better response for the same set of parameters compared to the NOT gate. Finally, for a lower value of $T_2^*$ a higher value of ac magnetic field is required for high fidelity gate operation.

\section{Toffoli gate}\label{Three_Qubit}
\begin{figure}[htbp]
\centering
\includegraphics[scale=0.4]{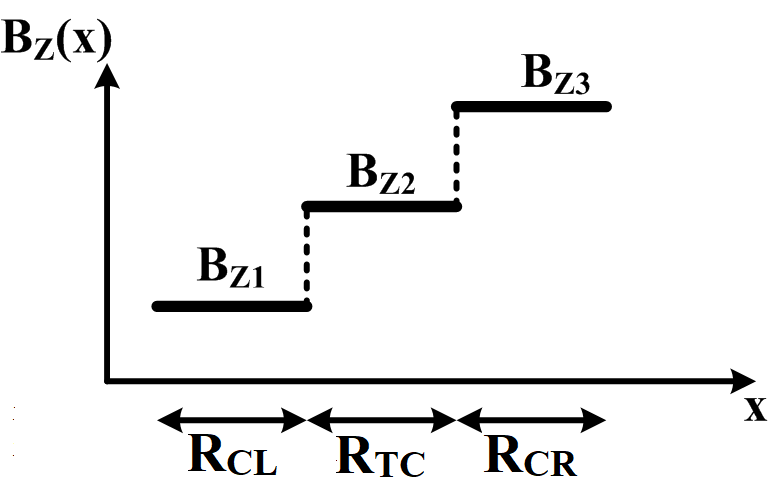}
\caption{Magnetic field profile for three-qubit operation. $R_{CL}$ corresponds to the region of left control qubit, $R_{TC}$ corresponds to the central target qubit, and $R_{CR}$ corresponds to the region of right control qubit.}
\label{Mag_Pro_Three_Spin}
\end{figure}

A system of three spin qubits can be used to implement controlled-controlled NOT (CCNOT) gate or Toffoli gate, as illustrated in Fig.~\ref{Mag_Pro_Three_Spin}~\cite{gullans2019protocol}. The Toffoli gate uses two control qubits and one target qubit. Based on the states of the control qubits, the target qubit is either flipped or left unchanged.  The complete Hamiltonian for three qubits Toffoli system is given by Eq.~\ref{Hamiltonian_of_three_spin_eqaution}, where $J_{12}$ gives exchange interaction between electrons in region $R_{CL}$-$R_{TC}$ and the $J_{23}$ is the exchange interaction for regions $R_{TC}$-$R_{CR}$ (cf. Fig.~\ref{Mag_Pro_Three_Spin}). The middle qubit acts as the target qubit while the other two act as control qubits.


\begin{equation} \label{Hamiltonian_of_three_spin_eqaution}
\begin{split}
 H(t) & = J_{12}(\sigma_{1x}\sigma_{2x}+\sigma_{1y}\sigma_{2y}+\sigma_{1z}\sigma_{2z})\\
 &   +J_{23}(\sigma_{2x}\sigma_{3x}+\sigma_{2y}\sigma_{3y}+\sigma_{2z}\sigma_{3z})\\
 &   +g\mu_{B}B_{1z}\sigma_{1z}+g\mu_{B}B_{2z}\sigma_{2z}+g\mu_{B}B_{3z}\sigma_{3z}\\
 &   -g\mu_{B}B_{ac}(coswt(\sigma_{1x}+\sigma_{2x}+\sigma_{3x})\\
   &   -\sin(wt)(\sigma_{1y}+\sigma_{2y}+\sigma_{3y}))
\end{split}
\end{equation}
The static magnetic fields $B_{z1}$, $B_{z2}$, $B_{z3}$ are used to create a magnetic field gradient ($\delta B_{z}$) where we assume that the magnetic field is constant along the length of an individual electron wave function. $B_{ac}$ is the ac magnetic field that is used to cause state transitions at a particular frequency. When both control qubits are in the state $\ket{\uparrow}$, the state of the target qubit is flipped; otherwise, it remains unchanged. This happens because the frequency of $B_{ac}$ correspond to energy gap $\ket{\uparrow_{CL}\uparrow_{TC}\uparrow_{CR}}\leftrightarrow\ket{\uparrow_{CL}\downarrow_{TC}\uparrow_{CR}}$.  We also assume that ($B_{z3}-B_{z2}=\delta B_{z32}, B_{z2}-B_{z1}=\delta B_{z21}, \delta B_{z32}=\delta B_{z21} $) and $J_{12}$=$J_{23}=J$.

\begin{figure}[htbp]
\centering
\includegraphics[width=90mm]{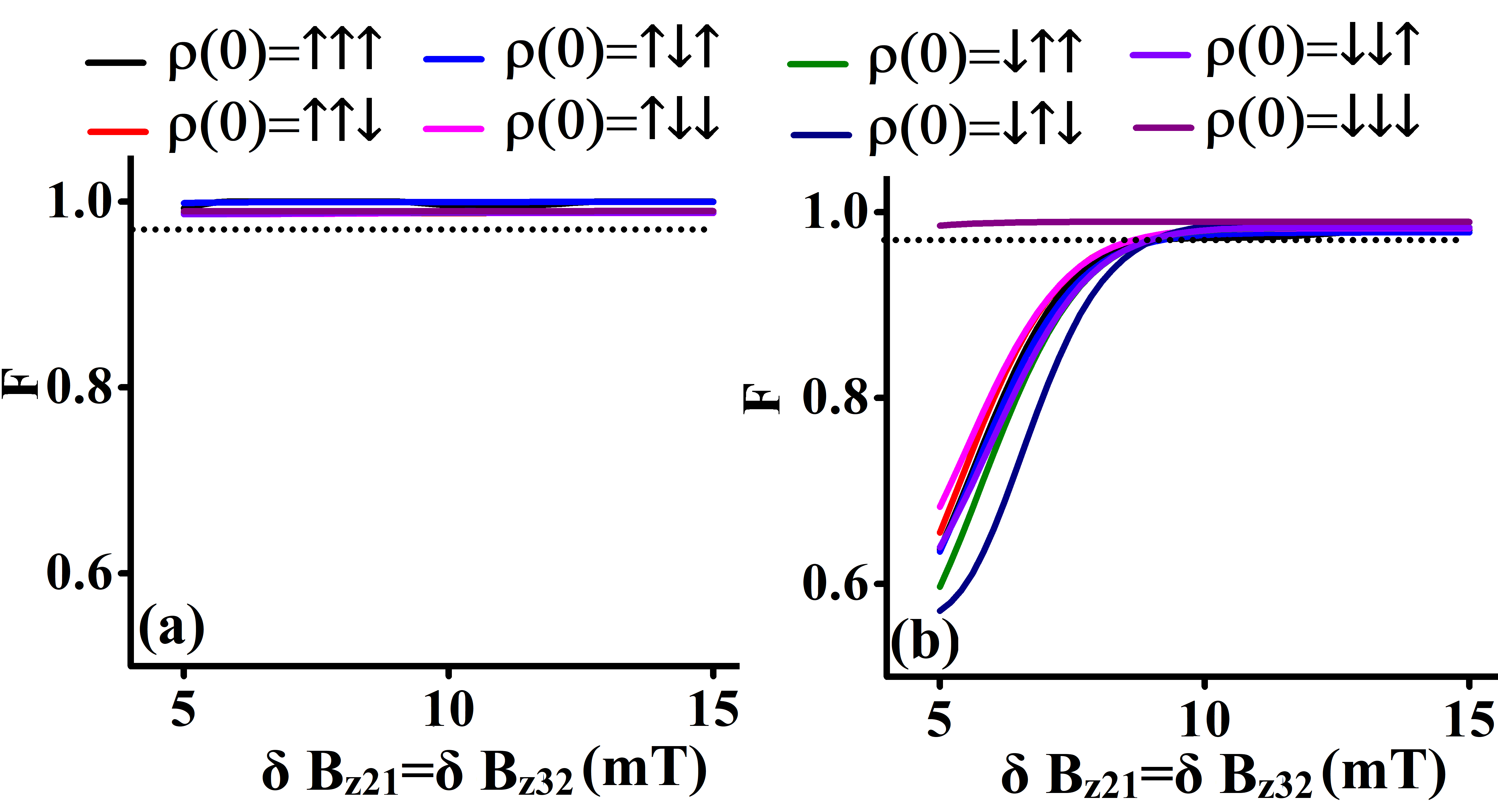}

\caption{(a) Fidelity vs magnetic gradient ($\delta B_{z}$) for $B_{ac}$ = 0.1mT, $B_{z1}$ = 40mT, and $J = 2.4MHz$ when no noise is considered in the system for all eight possible basis state. (b) Fidelity vs magnetic gradient ($\delta B_{z}$) for $B_{ac}$ = 0.1mT, $B_{z1}$ = 40mT, and $J = 2.4MHz$ when noise is considered in the system for all eight possible basis state. The dotted line corresponds to $F \geq$ 0.98 }
\label{Toffoli_low}
\end{figure}

The high fidelity Toffoli gate operation is demonstrated in Fig.~\ref{Toffoli_low}. In our analysis, we find that a high $B_{ac}$ gives a high-fidelity Toffoli operation where the ac field frequency corresponds to the energy gap between state$\ket{\uparrow_{CL}\uparrow_{TC}\uparrow_{CR}}\leftrightarrow\ket{\uparrow_{CL}\downarrow_{TC}\uparrow_{CR}}$. The gate fidelity is observed to be high ($\geq$ 98\%) for cases when $\rho(0)=\ket{\uparrow_{CL}\uparrow_{TC}\uparrow_{CR}}$ or when  $\rho(0)=\ket{\uparrow_{CL}\downarrow_{TC}\uparrow_{CR}}$. However, in all the other cases, fidelity will in fact reduce with a high ac field. For example: when $\rho(0)=\ket{\downarrow_{CL}\uparrow_{TC}\uparrow_{CR}}$, the state should remain unchanged, however due to high $B_{ac}$ we see oscillation of small magnitude which results in F$\leq$ 0.98. On the other hand, if the $B_{ac}$ is small, we find that the resonant state shows F$\leq$ 0.98. So it becomes a challenge to identify the $B_{ac}$ which shows high fidelity for all the basis states. From Fig.~\ref{Toffoli_low}, we find that when $B_{z1}$=40 mT, $J=2.4MHz$, $B_{ac}=0.1mT$ and $\delta B_z \geq$ 12mT, we obtain $F\geq0.98$.

From a practical standpoint, it is difficult to achieve gradients of more than a few mT/nm through micromagnets\cite{yoneda2015robust}. Therefore, practical Toffoli implementation may not possible at the high  magnetic field gradients (few T/nm) where phononic noise becomes a dominant source of decoherence. Owing to this, high-fidelity universal gate implementation is limited by multi-qubit operation rather than the single qubit based gates. Putting together the single qubit gates and Toffoli gate operation, we conclude that $B_{ac}$= 0.1 $mT$ is ideal for high-fidelity universal gate implementation.

\section{Conclusion}

In this work, we have presented a scheme for implementing high-fidelity universal quantum gates based on quantum dots in presence of hyperfine interaction noise and charge noise due to phonons.
We estimate the parameter range of static and ac magnetic fields corresponding to high-fidelity ($> 98 \%$ ) gate operations in noisy. 
In addition, we find that the parameter range of the gate operations is limited by the initial qubit state having the least energy in high static magnetic field regime. On the other hand, in low static magnetic field regime, the range is limited by initial qubit state having the least energy gap with respect to all other states. We also examine the interplay between the ac magnetic field and decoherence time in single qubit gates that gives us interesting insights about the magnetic field value to be used in a gate operation. We also identify a common parameter regime for Hadamard and Toffoli gate implementations. In future, we plan to devise noise cancellation techniques based on dynamical decoupling pulse sequences and gradient ascent pulse engineering (GRAPE) technique that can further improve the gate fidelity and possibly increase the operating range of these parameters and make the quantum gates more resilient thus reducing the error correction overhead. The framework presented in this work can be extended to a larger system (i.e. a multi-qubit quantum processor) to execute various gate operations and would be helpful in understanding the execution of quantum algorithms using quantum dot-based spin qubits. 

\vskip6pt



\bibliographystyle{IEEEtran}
\bibliography{IEEEabrv,Bibliography}


\vfill

\end{document}